# Phonon Localization in Heat Conduction

M. N. Luckyanova[1], J. Mendoza[1], H. Lu[2†], S. Huang[3], J. Zhou[1], M. Li[1], B. J. Kirby[4], A. J. Grutter[4], A. A. Puretzky[5], M. Dresselhaus[3,6], A. Gossard[2], and G. Chen[1*]

**Affiliations:**

[1]Department of Mechanical Engineering, Massachusetts Institute of Technology, Cambridge, MA 02139, USA

[2]Materials Department, University of California, Santa Barbara, CA 93106, USA

[3]Department of Electrical Engineering and Computer Science, Massachusetts Institute of Technology, Cambridge, MA 02139, USA

[4]Center for Neutron Research, National Institute of Standards and Technology, Gaithersburg, MD 20899, USA

[5]Center for Nanophase Materials Science, Oak Ridge National Laboratory, Oak Ridge, TN 37831, USA

[6]Department of Physics, Massachusetts Institute of Technology, Cambridge, MA 02139, USA

*Correspondence to:  gchen2@mit.edu
†Current address: College of Engineering and Applied Sciences, Nanjing University, Nanjing, 210093, China

Departures in phonon heat conduction from diffusion have been extensively observed in nanostructures through their thermal conductivity reduction and largely explained with classical size effects neglecting phonon's wave nature. Here, we report localization-behavior in phonon heat conduction due to multiple scattering and interference of phonon waves, observed through measurements of the thermal conductivities of GaAs/AlAs superlattices with ErAs nanodots randomly distributed at the interfaces. Near room temperature, the measured thermal conductivities increased with increasing number of SL periods and eventually saturated, indicating a transition from ballistic-to-diffusive transport. At low temperatures, the thermal conductivities of the samples with ErAs dots first increased and then decreased with an increasing number of periods, signaling phonon wave localization. This Anderson localization behavior is also validated via atomistic Green's function simulations.  The observation of phonon localization in heat conduction is surprising due to the broadband nature of thermal transport. This discovery suggests a new path forward for engineering phonon thermal transport.

**Main Text:**
**Introduction**
The phonons responsible for heat conduction in most dielectrics and semiconductors have short wavelengths. Although studies on phonon heat conduction in nanostructures over the past two decades have demonstrated the presence of strong size effects, most experimental observations of the departure from bulk behavior can be explained without invoking the wave nature of phonons[1]. Instead, the classical size effects happen when the phonon mean free paths (MFPs) are longer than the characteristic sizes of the structures, and these effects are important for a wide range of applications including thermoelectric energy conversion and microelectronic thermal management[2–6]. The potential of engineering phonon heat conduction via wave effects like bandgap formation[7–9], solitons[10], or localization[11,12] has been suggested before, but conclusive experiments have been lacking until recent demonstrations of coherent phonon heat conduction[13,14] in superlattices (SLs). Previous simulations and recent experiments revealed that, in SLs, most phonons responsible for heat conduction have mid- to long wavelengths because short wavelength phonons are strongly scattered by atomic mixing at individual interfaces[13,15–17]. These long wavelength phonons maintain their phases while propagating through the entire thickness of the SL. If such mid- to long wavelength phonons can be scattered, the thermal conductivity of the SLs can be further reduced. In this paper, we demonstrate the ability to control these mid-wavelength phonons by placing nanoscale dots at the interfaces of the SLs, leading to a further reduction of SL thermal conductivity by over a factor of two. Furthermore, our experiments and simulations demonstrate that these nanodots cause phonon localization over a broad frequency range, establishing a new paradigm for engineering phonon heat conduction in solids.

In the SL system, high-frequency phonons are primarily scattered by interface roughness and atomic mixing. Since the length scale of interface roughness is usually too small to effectively scatter long-wavelength phonons, they are instead dominantly scattered through anharmonic processes[13,15–17]. When the MFPs of these low-frequency phonons are longer than the entire sample length, they are scattered at the boundaries of the SLs rather than at the internal interfaces between layers. These propagating phonons represent eigenstates of the SL rather than of the parent material. Such coherent phonon transport has been previously observed in GaAs/AlAs SLs where the period thickness was held constant while the number of periods, and thus the overall sample thicknesses, was varied[13]. The thermal conductivity of these SLs depends on the total thickness $L$ of the SL. This effect was especially pronounced at lower temperatures, where long-wavelength phonons dominate thermal transport. By introducing scatterers with sizes comparable to the wavelengths of these phonons, further reduction in the thermal conductivity is expected. Furthermore, it was predicted that scattering centers randomly distributed at the interfaces of SLs may lead to photon localization[18]. It will be especially interesting to see if coherent phonons in the SL can be localized in a similar configuration.

**Experimental Evidence of Localization**
We fabricated three sets of seven SLs, with 4, 8, 12, 16, 100, 200, and 300 periods nominally comprising a 3 nm GaAs layer and a 3 nm AlAs layer on a semi-insulating GaAs (001) substrates (see Methods for details). The three sets of SLs are distinguished by the content of ErAs dots with an approximate diameter of 3 nm grown on top of the GaAs layers, at the interfaces with AlAs: (1) a reference set without dots at the interfaces (hereon referred to as the reference set), (2) a set with 8% (by area) interface surface coverage, leading to a spacing of about 9.5 nm between islands and (3) a set with 25% interface coverage, with an approximate inter-dot spacing of 5.5 nm. A schematic diagram of the SLs as well as TEM and HRTEM images of the samples are shown in Fig. 1 (further images are shown in Extended Data Figures 1 and 2). The images confirm the quality and consistency of the SLs throughout the whole thickness, and the random distribution of the dots at the interfaces. The significant areal coverage for the samples with 25% ErAs interfacial nanodot concentration leads to a slight variation in the surface planarity of the SLs that propagates through the SLs and manifests in a slight thickness variation at the surface, as can be seen in Extended Data Figure 2f, while the 8% areal coverage sample does not show a discernible variation from the bottom to the top of the thickest sample. The SL structure and interface quality were further confirmed with polarized neutron reflectometry (Extended Data Figure 3), which showed a period of 6.14 nm in the SL with 8% ErAs concentration and 6.2 nm in the SL with 25% ErAs concentration.

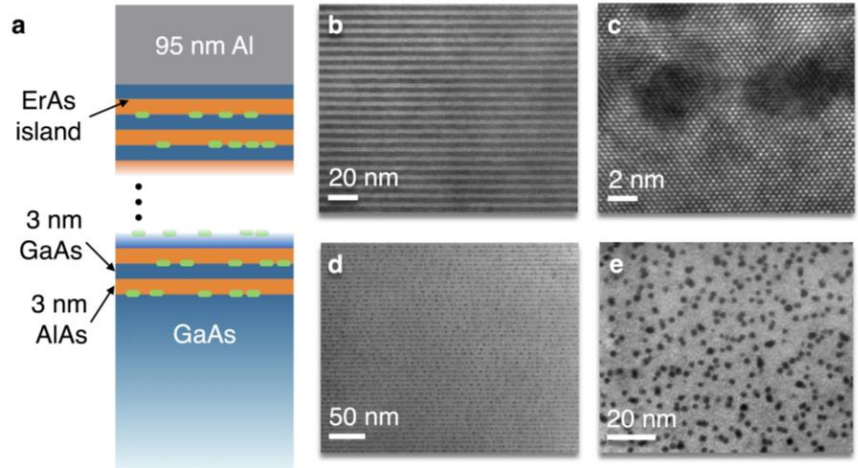

**Fig. 1 a,** Illustration of the SL samples. All samples have a fixed period thickness (3 nm of GaAs and 3 nm of AlAs) with varying numbers of periods. Three sample sets are distinguished by a varying ErAs dot content at the GaAs-AlAs interfaces, (1) no ErAs, (2) 8% areal coverage with dots, and (3) 25% areal coverage with dots; **b,** a cross-sectional TEM of the SL with no ErAs dots at the interfaces, **c,** a high-resolution TEM of the ErAs dots; **d,** a cross-sectional and **e,** plan-view TEM of the sample with 8% of the interfaces covered with ErAs dots.

The thermal conductivities of the SLs as a function of temperature, measured with time-domain thermoreflectance (TDTR), as described in Ref. 18, are shown in Figs. 2a-2f (representative data are shown in Extended Data Figures 4 and 5). For all the SLs, the thermal conductivity rapidly increases with temperature up to approximately 100 K, after which the thermal conductivity roughly plateaus. For small numbers of periods, the thermal conductivity also increases with an increasing number of periods, indicating the presence of coherent phonon thermal transport (Extended Data Figure 6).

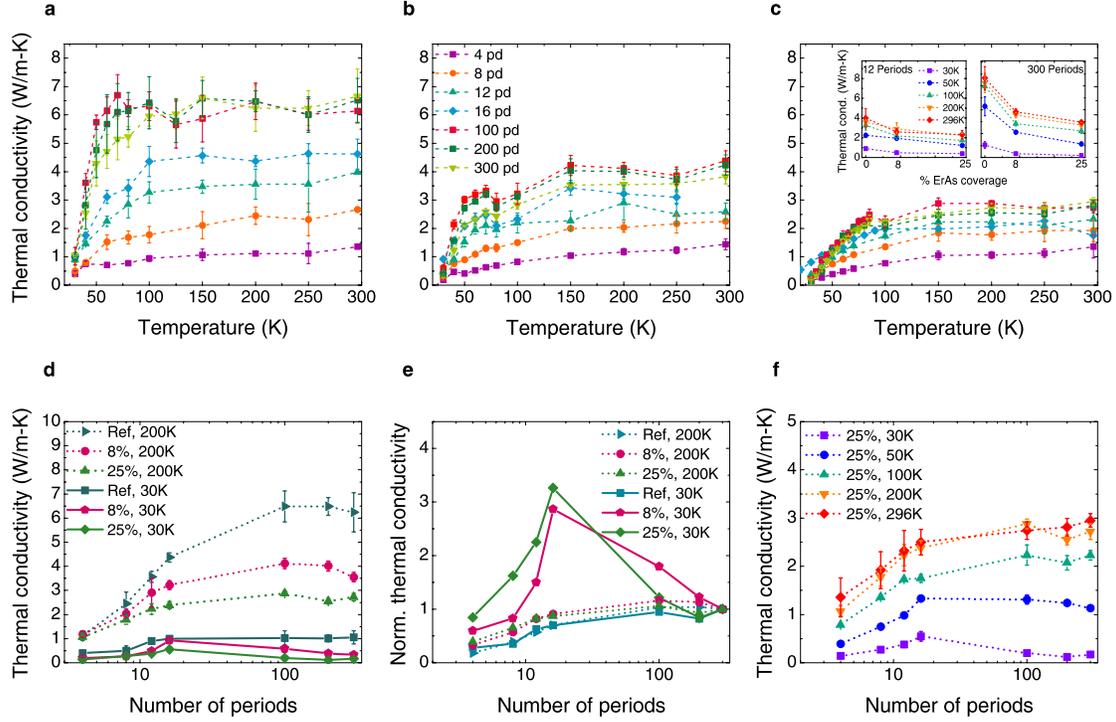

**Fig. 2** Thermal conductivity as a function of temperature for the temperature range 30 K - 296 K for the three sets of samples: **a,** the reference samples, which have no ErAs at the interfaces, **b,** the samples where 8% of the GaAs-AlAs interfaces are covered in ErAs dots, and **c,** the samples where 25% of the GaAs-AlAs interfaces are covered with ErAs dots. **d,** At 200 K, the thermal conductivity first increases with the number of periods and then saturates, suggesting that some phonon traverse the SLs coherently. At 30 K, the thermal conductivity behaves similarly in the reference sample, but in the samples with ErAs dots at the interfaces, the thermal conductivity decreases after reaching a peak at a small number of periods. **e,** When the thermal conductivities in **d** are normalized to the thermal conductivity of the longest sample, the anomalous low-temperature trend for samples with ErAs dots is even more pronounced. **f,** As the temperature is increased, the thermal conductivity of the samples with 25% ErAs dots begins to match the trend seen in the reference sample, a uniform increase of thermal conductivity with increasing number of periods. Error bars in the data represent +/- one standard deviation.

The addition of ErAs nanodots decreases the overall thermal conductivity, with a greater coverage of ErAs leading to a greater decrease in the overall thermal conductivity, as seen in the insets in Fig. 2c. High-frequency phonons are scattered by interface roughness and atomistic mixing. The addition of ErAs dots has the additional effect of scattering phonons over a wide frequency range as will be seen later, decreasing the range of coherent phonons that propagate through the entire SL without undergoing scattering. As more dots are added, this range grows. This is indicated by the decreasing thermal conductivity with increasing ErAs content and also, notably, by the greater overall decrease in thermal conductivity for the samples with more SL periods (insets in Fig. 2c), as compared to those with a small number of periods because in the shorter SLs, the boundary between the SL and the substrate plays a similar role in reducing the effective mean free paths. This latter effect bolsters the results of our previous work on coherent phonon heat conduction in SLs[13]. By adding scatterers at the SL interfaces, we are able to reduce the role of coherent phonon transport in SLs and to decrease the SL bulk thermal conductivity by a factor of two. We should also mention that the use of nanodots to increase the scattering of intermediate wavelength phonons has been studied before both theoretically and experimentally[19–22]. In particular, Pernot et al.[22] compared the measured thermal conductivities of Ge-dot based Si/Ge SLs with the simulated thermal conductivities of Si/Ge SLs with no dots and with perfect interfaces[22]. This comparison suggested that the Ge-dots based SLs can have 5x lower thermal conductivity than SLs without nanodots. The major differences between our study from the Pernot et al.[22] study and other previous studies are that (1) GaAs/AlAs SLs with and without ErAs dots have identical crystal structures, allowing for a direct comparison of experimental data, and (2) whereas previous studies focused on thermally thick SLs with different period thicknesses, we fixed the thickness of the SL periods while varying the total numbers of periods of the SLs. Our approach allows us to discover unique wave effects as we discuss below.

The dependence of the thermal conductivity on the total length of the SLs, $L$, reveals a very interesting trend (Fig. 2d). At temperatures above 100 K, the samples with a small number of periods (4, 8, 12, and 16 periods) uniformly show an increasing thermal conductivity with an increasing number of periods, which signifies that transport is primarily ballistic. With an increasing number of periods, the thermal conductivities of the SLs eventually saturate to a value that is much lower than either bulk parent material (45 W/m-K for GaAs and 90 W/m-K for AlAs[23]) and, in fact, the saturation values are lower than that for a corresponding bulk homogenous alloy (10 W/m-K). Extensive past modeling and experimental studies on the thermal conductivity of SLs, together with recent first-principles simulations, can explain these experimental observations well. This lower-than-bulk effective thermal conductivity is mainly due to the scattering of short wavelength phonons by the internal interface roughness of the SLs. However, long wavelength phonons are not effectively scattered by this interface roughness. They traverse the short-periods SLs ballistically while maintaining their phase, leading to an increasing thermal conductivity with an increasing number of periods. At a large number of periods, these coherent, long wavelength phonons are eventually scattered by inelastic phonon-phonon processes. Thus, on a coarse-grained level and in the large number of periods limit, both long and short wavelength phonons can be viewed as undergoing diffusive transport, even though high frequency phonons are dominantly scattered at individual interfaces and long wavelength phonons are scattered via phonon-phonon processes after traversing many interfaces, leading to the experimentally observed saturation of thermal conductivity with increasing number of periods.

At lower temperatures, SLs having ErAs-induced disorder at the interfaces show a different and surprising trend. Below 60 K, the SLs with 25% ErAs coverage at the GaAs-AlAs interfaces and fewer periods (12 and 16 periods) have a higher thermal conductivity than the longest SLs, as seen in Figs. 2c and d (low-temperature data where this transition is clearly visible are shown in Extended Data Figure 7). This transition also happens in the samples with 8% ErAs coverage at the GaAs-AlAs interfaces, but is absent from the reference samples. To show this more clearly, we normalize the measured thermal conductivities for all SL periods to the 300 period SL value for each sample. The resulting normalized thermal conductivity is shown in Fig. 2e and a peak clearly exists in both the 8% and 25% samples at 30K. The observation of this trend in samples with two different concentrations of ErAs nanodots at the interfaces suggests that, rather than being caused by small variations in the arrangement of dots in the 25% coverage sample, a new heat conduction mechanism, which reduces the transport of long-wavelength, low-frequency (THz range) phonons, is unfolding with an increasing sample thickness. This unexpected trend strongly points to the presence of phonon localization in these samples. Localization effects in phonon transport are expected to unfold over an increasing sample thickness, as certain phonon modes make the transition from being propagating to non-propagating modes due to the destructive interference from multiple elastic scattering events[24]. Localization is characterized by a phonon frequency-dependent localization length, $\xi$. Phonon transmission decays exponentially for localized modes as $e^{-L/\xi}$, where $L$ is the total thickness of the SL. When $L$ is small compared to $\xi$, phonon transport is in the ballistic regime and the thermal conductivity increases with the total thickness. As the total thickness of the SLs increases, some phonons are localized and do not contribute to heat conduction anymore, leading to a decreasing thermal conductivity with increasing length, $L$. This transition from ballistic transport to localized transport is evident by the presence of a maximum in the measured thermal conductivity as shown in Figs. 2d-f, and will be supported by simulations presented later.

We do not observe the localization of heat conduction at room temperature because with increasing temperature, a broader range of phonon frequencies contributes to the thermal conductivity. Higher frequency phonons experience diffuse interface scattering and phase-breaking phonon-phonon scattering, leading to a trend of saturating thermal conductivity with increasing number of periods (Figs. 2e and 2f). This is correlated with the ballistic-to-diffusive transition typically observed in nanostructures. At lower temperatures, the phonon population shifts more to lower frequency phonons so phonon-phonon scattering becomes less effective, making the contributions from localized phonons observable. This ballistic-to-localized phonon transport behavior has never been observed before.

**Simulations Supporting Localization**
We carried out simulations to further explain the experimental observations. We use atomistic Green's function simulations to compute phonon transmission across SLs with (1) perfect interfaces, (2) atomic mixing (roughness) at interfaces, and (3) roughness and nanodots at each interface, following established methods with accelerated algorithms[13,25] (see Methods). Due to computational limitations, we rescaled the lateral size of the system so that the cross-sectional area was 1.68 nm x 1.68 nm ($3a$ x $3a$, where $a$ is a lattice constant) with periodic boundary conditions in the lateral direction. The interface roughness comprises random mixing of Ga and Al atoms in one crystal lattice on each side of the interfaces and the nanodots were spheres with diameter of two lattice constants, or 1.12 nm, with their centers positioned randomly at both the GaAs-AlAs and AlAs-GaAs interfaces. The results presented in fig. 3a-e represent the averages of an ensemble of 20 random distributions of nanodots. Figure 3a shows the calculated transmission functions for SLs with only atomic-mixing-caused roughness at the interfaces and for SLs with both roughness and nanodots at the interfaces. As the number of periods is increased, the transmission function decreases in both of these scenarios, but the decrease is much more drastic in the samples with both roughness and nanodots. The transmittance as a function of frequency for these cases further highlights this effect (Fig. 3b). At frequencies around 2 THz, there is a significant dip in the transmission functions with increasing numbers of periods in the SLs with roughness and nanodots. From these figures, we observe that phonon transmittance is significantly reduced in samples with nanodots for frequencies near the zone edges (near 1.5 THz), consistent with the established picture that it is easier to localize at band edges[18,24] (a calculated dispersion relationship is shown in Extended Data Figure 8). In Fig. 3c we show the transmission function as a number of periods for normal incidence phonon with a frequency of 1.65 THz, which shows an exponential decay in the transmission of phonons of this frequency in the SLs with roughness and nanodots. Figures 3a and 3b also show that nanodots localize phonons over a wide frequency range.

We follow the formalism in MacKinnon *et al.*[26] to compute the localization lengths in the 300-period samples at different frequencies (Fig. 3d) in the SL structures with interface roughness and nanodots. Also shown in Fig. 3d are inelastic phonon mean free paths due to phonon-phonon scattering obtained from first-principles simulations[13] at 30 K and 300 K. At 30 K, the inelastic mean free path on average is much longer than the localization length so that phonons maintain their phases during transport and localization can occur. In Fig. 3e, we show the contributions of each phonon frequency to the overall thermal conductivity for the three 100 period samples in Fig. 3c with a thermal conductivity accumulation plot. The accumulated thermal conductivity of the SL with nanodots has a flat region between 1.5 and 2.8 THz. In this range, most phonons are localized and do not contribute to thermal conductivity. Using the transmission functions, we calculate the thermal conductivity as a function of number of periods, neglecting the effects of anharmonic scattering which is much longer than the total SL thickness. Figure 3f shows the normalized thermal conductivity, while the inset in Fig. 3f shows the computed thermal conductivity at 30 K of SLs with interface roughness and SLs with both interface roughness and nanodots as a function of the number of periods . These values are normalized to the value at 300 periods. We note that the predicted thermal conductivity value and peak location vs. period is in the same range as that observed in the experiments (Fig. 2e). The consistency of both the existence of the peak, and its location as a function of number of periods between he simulations and experiments further supports our explanation that the experimental observation is due to phonon localization.

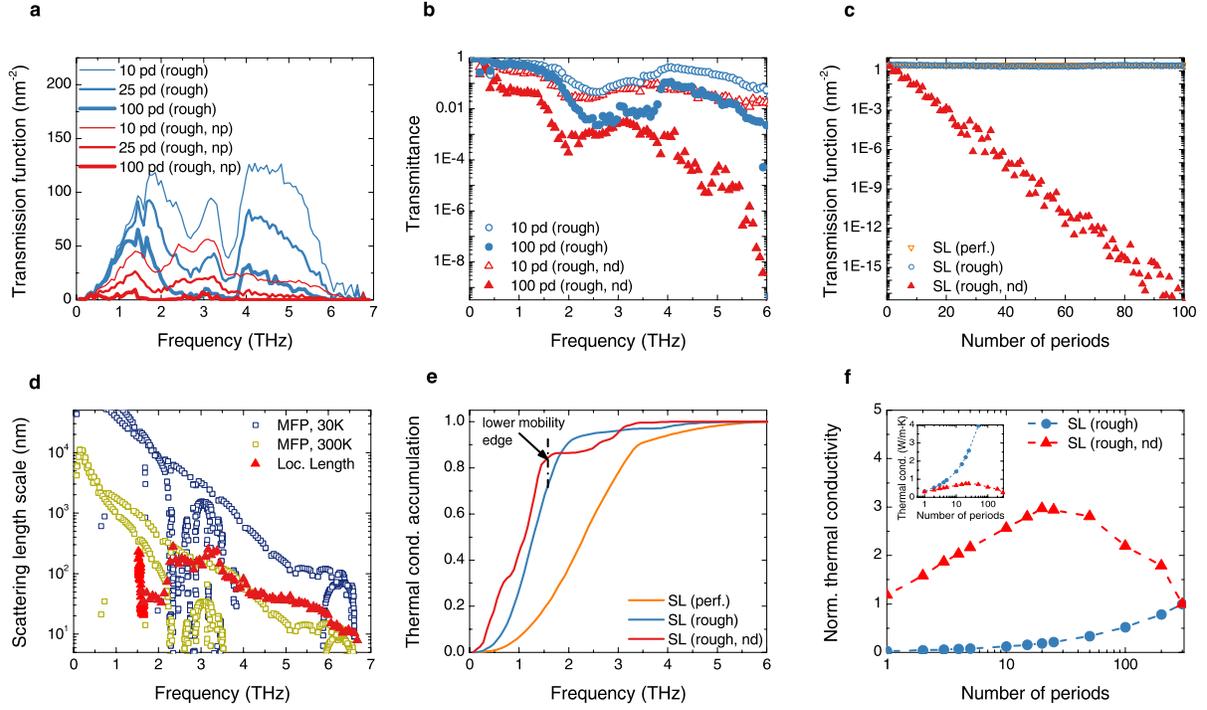

**Fig. 3** Frequency of the dependence of the **a,** transmission functions summed over all angles with increasing number of periods and **b,** transmittances of phonons in SLs of different frequencies with only roughness and with both roughness and nanodots. **c,** Transmission function as a function of the number of SL periods for SLs with perfect interfaces, roughness at the interfaces, and roughness and nanodots at the interfaces for 1.65 THz frequency phonons at normal incidence. **d,** Inelastic MFPs of a perfect SL at 30 K and 300 K and localization length in a SL with both roughness and nanodots at the interfaces. **e,** Thermal conductivity accumulation for perfect SLs, SLs with roughness at the interfaces, and SLs with both roughness and nanodots at the interfaces at 30 K. **f,** Normalized thermal conductivity as a function of number of SL periods for rough SLs with and without ErAs nanodots at the interfaces at 30 K (inset shows calculated thermal conductivity values).

## Discussion

The localization of waves in general has been studied extensively since the pioneering work of Anderson[18,24,27,28]. It has been found that in disordered 1D and 2D media, all finite frequency phonons are localized, while in 3D, a mobility edge separating long frequency extended states and high frequency localized states exists[24,29]. However, heat conduction is a broadband phenomenon and no experiments have clearly shown the impact of localization. In bulk dielectric and semiconductor crystals, where phonon transport dominates heat conduction, the wave vectors, $k$, at the band edges are high. In accordance with the modified Ioffe-Regel criterion for strong wave localization in three dimensions, $l \times k \leq 1$[30], localized phonons in a bulk material have MFPs, $l$, on the order of angstroms. Since phonons contributing to the thermal conductivity in bulk crystals have much longer MFPs, the Ioffe-Regel condition is difficult to meet and observations of localization have been elusive. Furthermore, not all phonons satisfying the Ioffe-Regel criterion are localized[31]. A minimum in the thermal conductivities of some SLs was used by Venkatasubramanian as a sign of localization[32], but later works showed that these minima were the result of the competition between coherent and incoherent transport in SLs[14,33]. The plateaus in the thermal conductivities of glasses and aggregates at low temperatures were thought of as the combined effects of localization and diffusion, although simulations show that localization was not dominant[34]. In these materials, however, MFPs are on the order of tens of angstroms, much shorter than in typical crystalline materials. Computational studies of thermal transport in BN with isotopic impurities, where the mass mismatch was small, found that localization is unobservable in thermal transport properties[35]. A few publications have shown traces of phonon localization in thermal transport simulations, including molecular-dynamics simulations in SLs with randomized period thicknesses[12] and continuum-mechanics based simulations of cross-section modulated core-shell nanowires SLs[11]. These simulations more closely model low-dimensional structures, for which localization is easier to observe.

We can observe the consequences of phonon localization on thermal conduction through ErAs-dot doped GaAs/AlAs SLs because of several factors. First, natural interface mixing in GaAs/AlAs SLs scatters high-frequency phonons and reduce their contribution to thermal conductivity (Fig. 3a and 3b). Second, zone folding in SLs reduces the wavevector at the BZ boundary by approximately an order of magnitude, from the lattice constant of bulk crystals (~5 Å) to the 6 nm lattice constant of the SL. The reduced wavevector allows longer MFP phonons to satisfy the modified Ioffe-Regel criterion. In fact, the calculated localization length in the 300 period SLs with nanodots diverges at 1.58 THz, (Fig. 3d) which we identify as the lower mobility edge. Furthermore, Fig. 3d also shows that the localization length jumps around 2.3 THz. This window of short localization lengths is consistent with theoretical prediction by Kirkpatrick on acoustic waves[29] and John on photon localization in SLs with lateral disorder[18]. Even above the second mobility edge at 2.3 THz, however, our simulated localization length is still shorter than the ineslastic scattering length, suggesting localization is still possible even at higher frequencies. Third, the diameter of the ErAs is approximately 3 nm, and on average their spacing are 9 nm for the SLs with 8% coverage and 5 nm for the SLs with 25% areal coverage. These are comparable to the phonon wavelengths at the boundaries of the folded zone, increasing the multiple scatterings of these phonons, leading to their localization.

At higher temperatures, some low frequency phonons can still be localized despite the absence of a localization signal in the measured thermal conductivity, as suggested by Fig. 3d, since these phonons have inelastic mean free paths longer than the localization length. We conducted a Raman scattering experiment to probe the first few bands of the folded band structure in the SLs. The measured phonon spectra of the first two bands of selected SLs at room temperature is shown in Extended Data Figure 9. The Raman experiment measures transverse phonons at the zone center rather than at the band edge[36] While the reference SLs show clear peaks at 0.75, 0.90, 1.65, and 1.80 THz, the ErAs covered SLs do not display strong signals at the latter two bands. For 25% ErAs coverage, the first band disappears in the 300 period SL. We interpret the disappearing bands as a result of wave cancellation that leads to localization. We also measured the pump modulation frequency dependence on the measured thermal conductivities of the SLs, and the experimental results are consistent with the localization picture we presented (Extended Data Figure 10, with further discussion in the Supplementary Information).

In summary, we have demonstrated the potential of engineering phonon waves to reduce the thermal conductivity in SLs. By using ErAs nanodots, we reduce the thermal conductivity of thick SLs by as much as a factor of two compared to samples without dots. At low-temperatures, we observed experimentally that the thermal conductivity of such SLs first increases and then decreases with the number of periods at low temperatures, consistent with the phonon localization effect. Our Green's function-based simulations further confirm this ballistic to localized transport transition. At higher temperatures, the thermal conductivity versus the number of SL period follows the ballistic to diffusion transition behavior. These observations open up new opportunities to engineer phonon thermal conductivity via wave effects.

## Methods

Sample Preparation

GaAs/AlAs superlattices (SLs) with ErAs nanoparticles are grown epitaxially[37]. All the samples were grown in a Veeco Gen III MBE system[‡]. Solid source materials Ga, Al, Er and a Veeco valved cracker for As, in addition, were used for the growths. Sub-monolayer (ML) ErAs deposition was inserted at the GaAs-AlAs interfaces and the two surface coverages of 8% and 25% were aimed by 0.32 ML and 1 ML ErAs deposition, respectively. Depositions of ErAs are described in MLs as if the ErAs had grown in a layer-by-layer growth mode (1 ML of ErAs = 2.87 Å of film thickness). Nanodots are formed due to self-assembly of the deposited atoms. The growth rates of GaAs and AlAs were set to be around 1 µm/hr, while the ErAs growth rate was kept at 0.02 ML/sec. The $As_2$ overpressure was maintained at $1.07 \times 10^{-3}$ Pa during the growth. All the samples were finished with a 3 nm GaAs layer to prevent possible oxidation of either AlAs or ErAs. ErAs has a lattice constant of 5.74 Å while that of GaAs is 5.6533 Å, a lattice constant mismatch of 1.53%. In general, we believe the growth rate calibration of ErAs can have a variation of a few percentages. And we estimate the surface coverage by dividing the ErAs layer thickness (from the growth rate calibration) by 4 monolayers (which was observed from previous TEM studies showing that most of the ErAs dots tend to form 4 ML tall). TEM images of the different sample configurations are shown in Fig. 1 of the main text and Extended Data Figure 1.

Polarized Neutron Reflectometry

The structure and composition of the 8-period SL samples are characterized by polarized neutron reflectometry (PNR), performed at the PBR beamline at the NIST Center for Neutron Research, with an in-plane guide field $\mu_0 H = 0.7T$ and at T = 4K (Extended Data Figure 3). Compared with the 8%-ErAs sample (pink curve in Extended Data Figure 3), the 25%-ErAs sample (green curve in Extended Data Figure 3) shows a shallower dip, indicating an increase of interface roughness upon the addition of the ErAs nanodots. The higher frequency oscillations observed for the 25% ErAs SL indicates a slight increase of the lattice period, which is quite reasonable since the lattice constant of ErAs exceeds that of GaAs by 1.53%. In order to understand the detailed structure and further information about the concentration of ErAs at the interfaces, we refine the PNR curves using a differential evolution algorithm by GenX[38], where the structure, roughness and ErAs concentration are extracted. The results are consistent with those observed with TEM analysis.

---

[‡] Certain commercial equipment, instruments, or materials are identified in this paper to foster understanding. Such identification does not imply recommendation or endorsement by the National Institute of Standards and Technology, nor does it imply that the materials or equipment identified are necessarily the best available for the purpose.

Time-domain Thermoreflectance
Briefly, a high-energy (~100 mW) laser pulse impinges on a metal opto-thermal transducer layer on the sample, exciting surface electrons which quickly thermalize, sending a heat pulse propagating through the metal and then through the sample, away from the surface. A time-delayed probe pulse measures the changing surface reflectivity with changing temperature. A multi-dimensional, multi-layer heat equation is fit to the resulting cooling curve, yielding the thermal conductivity of the SL layers[13]. The system used herein is described in further detail in a previous publication[39]. We deposited a 100 nm Al optothermal transducer layer on the SLs and a calibration sapphire substrate. We confirmed the thickness of the Al transducer layer by matching the TDTR-measured thermal conductivity value of the sapphire substrate with the known literature value. Low-temperature measurements were conducted by mounting the samples in a low-vacuum (~$10^{-3}$ Pa) cryostat. The transient reflectance of the surface of the samples was measured with a Si diode.

Each sample was probed at between three or five different locations. At each location, the samples were measured under four different pump modulation frequencies – 3, 6, 9, and 12 MHz, with three individual data traces collected for each modulation frequency. The three runs at each frequency were averaged, and the Fourier fitting analysis was performed on the resultant average curve at each modulation frequency for each location. Sample sets of fitting curves for the reference samples and the 25% ErAs coverage samples for different SL thicknesses and at different temperatures are shown in Extended Data Figures 3 and 4. Error bars in the plots in the main body of the paper and in the SI represent the standard deviation from the results of these fittings.

The data were fit to a four-layer model comprising the metal optothermal transducer, the interface between the metal and the SL, the SL, and a semi-infinite substrate. The data were fit for the interface conductance between the metal and the SL, and the SL thermal conductivity. The other required parameters for the fitting were taken from the literature. Since the volumetric fraction of ErAs is small compared to GaAs and AlAs (0.32 ML and 1 ML of ErAs per 10 ML of GaAs or AlAs for the 8% and 25% nanodot areal coverage samples, respectively), the heat capacity was taken to be the average of GaAs and AlAs. An initial guess for the interface conductance between the SL, capped by a 3 nm GaAs layer, and the Al optothermal transducer layer were taken from previous measurements of the interface conductance between a GaAs substrate and an Al layer deposited with an identical procedure.

Superlattice Dispersion and Dynamical Matrix:
The phonon dynamics are governed by the dynamical matrix, whose matrix elements, in real space, take the form $D_{ij} = \frac{\phi_{ij}}{\sqrt{m_i m_j}}$, where $\phi_{ij}$ represents the interatomic force constant (IFC) that couples the $i^{th}$ and $j^{th}$ atoms of mass $m_i$ and $m_j$, respectively. The bulk GaAs interatomic force constants are obtained from Density Functional Pertubation Theory (DFPT) [40] which utilizes a 16x16x16 Monkhorst-Pack $k$-point grid[41] and an energy cutoff of 80 Ry for the plane wave expansion. We used a 6x6x6 supercell for the simulations, which corresponds to a phonon Brillouin zone mesh size of 6x6x6. The interatomic force constants are obtained using the PWscf and PHonon packages within Quantum Espresso[42] using the psuedopotential from Bachelet-Hamann-Schlüter [43].

The GaAs/AlAs SL was constructed using bulk GaAs IFCs up to the fifth nearest neighbor and the masses of the SL structure. This method, known as the mass approximation, has successfully reproduced the AlAs phonon dispersion from GaAs IFCs and the GaAs phonon dispersion from AlAs IFCs [44]. Although this model does not describe the local strain effects from neither the interfaces nor the erbium nanodots, the dispersion of the phonon frequency with wave vector from Γ to X (Extended Data Figure 8) agrees well with the previously calculated dispersion for a 24 nm period GaAs/AlAs SL using semi-empirical force constants in the work of Luckyanova *et al.*[13]

Green's Function Simulation
The atomistic Green's function[25] approach models the heat transfer through a finite size device that is coupled to semi-infinite reservoirs on each end. The dynamical matrix of the entire system can be written as,

$$H = \begin{bmatrix} H_L & \tau_{LD} & 0 \\ \tau_{LD}^\dagger & H_D & \tau_{DR} \\ 0 & \tau_{DR}^\dagger & H_R \end{bmatrix} \quad \text{(S.1)}$$

where $H_L$, $H_R$, and $H_D$ are the dynamical matrices of the left reservoir, right reservoir, and device region, respectively. $\tau_{LD}$ is the dynamical matrix that couples the left reservoir to the device and $\tau_{DR}$ is the dynamical matrix that couples the device to the right reservoir. This formalism is only valid when the reservoirs are uncoupled from one another. For semi-infinite reservoirs, the dynamical matrix of equation (S.1) is infinitely large. To make this problem tractable, the interactions between the device and the reservoirs are encoded in self-energy terms $\Sigma_L = \tau_{LD}^\dagger g_L \tau_{LD}$ and $\Sigma_R = \tau_{RD}^\dagger g_R \tau_{RD}$ where $g_{L,R}$ are the surface Green's functions obtained from a real space decimation method [45]. The Green's function of the device region can now be computed as $G_D = (\omega^2 - H_D - \Sigma_L - \Sigma_R)^{-1}$ where $\omega^2$ is the square of the phonon eigenfrequency. Defining $\Gamma_{L,R} = i(\Sigma_{L,R} - \Sigma_{L,R}^\dagger)$, the transmission function for a given frequency $\omega$ and transverse wavevector $k$ can be written as

$$\Xi(\omega, k) = Tr[\Gamma_L G_D \Gamma_R G_D^\dagger], \quad \text{(S.2)}$$

noting that the frequency and wavevector arguments on the RHS are implicit. Defining $\Xi(\omega) = \frac{1}{N_k} \sum_k \Xi(\omega, k)$ as the normalized sum over $N_k$ points in the Brillouin zone, the thermal conductance at a given temperature $T$ can be expressed as

$$K(T) = \frac{1}{2\pi A_D} \int_0^\infty \hbar\omega \frac{\partial f(\omega, T)}{\partial T} \Xi(\omega) d\omega \tag{S.3}$$

where $A_D$ is the area of the device's cross section and $f(\omega, T)$ is the Bose-Einstein distribution function.

The computation of equation (S.3) only requires the subspace of Green's function matrix elements that connect the right reservoir and left reservoir. We shall denote the set of matrix elements of this subspace as $G_{1N}$. $G_{1N}$ is recursively computed from the Dyson's equation. Since $G_{1N}$ corresponds to the probability amplitude for a phonon to propagate across the entire device, the localization length $l_{loc}$ can be determined from[26]

$$\frac{1}{l_{loc}} = -\lim_{N \to \infty} \frac{\ln(\text{Tr}[|G_{1N}|^2])}{2Nl} \tag{S.4}$$

for a system of $N$ periods of length $l$. Due to computational limitations, localization lengths were extracted from an average of ten configurations of 300 period (~1700nm) devices, which also correspond to the thickest measurement samples.

*Modeling the Randomness:*
Since the atomistic Green's function method models harmonic systems, scattering comes from the breaking of translational symmetry. In this particular case, the suppression of phonon transport is due to the presence of interfacial roughness and ErAs nanoparticles. Within one unit cell on each side of the interface, coordinates corresponding to the Ga/Al sublattice are chosen at random. In the case of interface roughness, the Ga or Al atom is replaced with Al or Ga, respectively. In the case of randomly placed ErAs nanoparticles, the coordinate corresponds to the center of the impurity; consequently, all Ga and Al atoms within 0.5 nm of this point are replaced with Er atoms. Since the randomness introduces variance in the calculated thermal conductivity, a configurational average is necessary. The number of configurations, for a given sample length, is inversely proportional to its sample length to ensure constant computation cost for each data point in figures 3**e** and 3**f**. For our data, one configuration was used for the 100 period device regions, four configurations were used for the 25 period device region, and 100 configurations were used for the 1 period device region, etc.

First Principles Calculations
The phonon mean free path due to anharmonic phonon-phonon scattering is calculated by $\Lambda_{q\lambda} = v_{q\lambda}\tau_{q\lambda}$, where $v_{q\lambda}$ is the phonon group velocity and $\tau_{q\lambda}$ is the phonon relaxation time. The group velocity is obtained from the phonon dispersion, which is derived from the dynamical matrix set up by the harmonic force constants. The relaxation times are calculated using the lowest-order three-phonon scattering process via [46,47]

$$\frac{1}{\tau_{\mathbf{q}\lambda}^{ph-ph}} = \frac{1}{2h^2 N_{\mathbf{q}}} \cdot$$

$$\sum_{q_1\lambda_1,q_2\lambda_2} |V_{\mathbf{q}\lambda,\mathbf{q}_1\lambda_1,\mathbf{q}_2\lambda_2}|^2 \left\{ (n^0_{\mathbf{q}_1\lambda_1} + n^0_{\mathbf{q}_2\lambda_2} + 1) \cdot \begin{bmatrix} \delta(\omega_{\mathbf{q}_1\lambda_1} + \omega_{\mathbf{q}_2\lambda_2} - \omega_{\mathbf{q}\lambda})\delta_{\mathbf{q}+\mathbf{q}_1+\mathbf{q}_2,\mathbf{G}} \\ -\delta(\omega_{\mathbf{q}_1\lambda_1} + \omega_{\mathbf{q}_2\lambda_2} + \omega_{\mathbf{q}\lambda})\delta_{\mathbf{q}+\mathbf{q}_1+\mathbf{q}_2,\mathbf{G}} \end{bmatrix} + (n^0_{\mathbf{q}_2\lambda_2} - n^0_{\mathbf{q}_1\lambda_1}) \cdot \begin{bmatrix} \delta(\omega_{\mathbf{q}_1\lambda_1} - \omega_{\mathbf{q}_2\lambda_2} - \omega_{\mathbf{q}\lambda})\delta_{\mathbf{q}+\mathbf{q}_1+\mathbf{q}_2,\mathbf{G}} \\ -\delta(\omega_{\mathbf{q}_1\lambda_1} - \omega_{\mathbf{q}_2\lambda_2} + \omega_{\mathbf{q}\lambda})\delta_{\mathbf{q}+\mathbf{q}_1+\mathbf{q}_2,\mathbf{G}} \end{bmatrix} \right\} \quad (S.5)$$

where $V_{q,\lambda,q_1,\lambda_1,q_2,\lambda_2}$ are the three-phonon coupling matrix elements and depend on the third order force constants [46]. All the force constants are obtained by taking the average of those for pure GaAs and AlAs, which is a good approximation due to the small lattice mismatch between GaAs and AlAs and has previously been used for calculating the thermal conductivity of SLs [13]. The force constants for GaAs and AlAs are fitted, based on the first-principles data regarding forces acting on different atoms and their displacements in a large supercell (2x2x2 conventional unit cells, 64 atoms), with imposed translational and rotational invariances. The force-displacement data are computed using the density functional theory as implemented in the QUANTUM ESPRESSO package[42]. We use the norm-conserving pseudopotential with the Perdew and Zunger[48] local density approximation (LDA) for the exchange-correlation functional, a cutoff energy of 60 Ryd and a 16x16x16 **k**-mesh.

We have considered the pure GaAs/AlAs SL and the lattice constant is chosen to be the average (5.5722Å) of the calculated values in pure GaAs and AlAs. Due to the computational complexity, we have chosen the SL structure to have three conventional cells for each layer (GaAs or AlAs) in the z-direction (dimensions: 5.5722Å×5.5722Å×33.43Å). A 16x16x3 **q**-mesh is used for calculating the phonon relaxation times (Eq. S.5) and the convergence with respect to the mesh has been checked. We have also checked the phonon relaxation times of the SL with smaller period (dimension: 5.5722Å×5.5722Å×22.29Å) and found that the phonon mean free paths have small differences. Therefore our calculated results should be close to the experimental configuration, which has a slightly larger period.

Raman Scattering Experimental Configuration
Raman spectra for several samples are shown in Extended Data Figure 9. The low-frequency (<100 cm$^{-1}$) Raman spectra were obtained with a Horiba Jobin-Yvon T64000 triple-grating Raman spectrometer under a $Z(XX)\bar{Z}$ back-scattering configuration. The excitation laser was a frequency-doubled Nd:YAG laser with a wavelength of 532.1 nm, and the power incident on the sample was 7.2 mW. The acquisition time was 60 seconds to ensure a sufficient signal-to-noise ratio. A 100X objective lens with a numerical aperture (NA) of 0.95, a motorized XYZ stage, and three 1800 grooves/mm grating were used. Both Stokes and anti-Stokes Raman lines were collected and were used to improve the measurements of Raman shifts. In contrast to the low-frequency measurement, high-frequency (>100 cm$^{-1}$) spectra were measured with a Horiba Jobin-Yvon HR800 Raman spectrometer using an objective lens with an NA of 0.90, an incident excitation laser power on the sample of 1.2 mW, and a shorter accumulation time of 15 seconds. The other measurement configurations were the same.

**Acknowledgments:** We would like to thank Profs. Ping Sheng and Zhaoqing Zhang for discussion. We also thank A. A. Maznev, K. A. Nelson, E. N. Wang, S. Huberman, V. Chiloyan, and L. Zeng. Authors acknowledge TEM Analysis Services Lab for the TEM images. Work at MIT was supported by the Solid State Solar-Thermal Energy Conversion Center (S$^3$TEC), an Energy Frontier Research Center funded by the U.S. Department of Energy, Office of Science, Office of Basic Energy Sciences under Award DE-SC0001299. Work at UCSB was supported by the Center for Energy Efficient Materials (CEEM), an Energy Frontier Research Center funded by the U.S. Department of Energy, Office of Science, Office of Basic Energy Sciences under Award DE-SC0001009. The low-frequency Raman spectra measurement was conducted at the Center for Nanophase Materials Sciences, which is sponsored at Oak Ridge National Laboratory by the Scientific User Facilities Division, Office of Basic Energy Sciences, U.S. Department of Energy. We acknowledge the support of the National Institute of Standards and Technology, U.S. Department of Commerce, in providing the neutron research facilities used in this work.

# Supplementary Information for Manuscript:
# Phonon Localization in Heat Conduction

M. N. Luckyanova[1], J. Mendoza[1], H. Lu[2†], S. Huang[3], J. Zhou[1], M. Li[1], B. J. Kirby[4], A. J. Grutter[4], A. A. Puretzky[5], M. Dresselhaus[3,6], A. Gossard[2], and G. Chen[1*]

**Note on Raman Results**

It is interesting to note from Extended Data Figure 9 that at high frequencies, the Raman signal does not change with an increasing number of periods as drastically as it does at low frequencies. We interpret this as a result of the dominance of the inelastic scattering at high frequencies at room temperature, while the weakening or completely disappearance in the Raman signal at low frequencies with increasing number of periods is a result of localization, since these low frequency phonons have mean free paths (MFPs) comparable to or longer than the localization length according to Fig. 3d.

**Note on the Modulation Frequency Dependence of Thermal Conductivity**

The pump modulation frequency dependence of the thermal conductivity provides further evidence of the significant MFP reduction in SLs with ErAs nanodots. While *k* decreases with increasing modulation frequency in the 300 period reference SLs, as seen in Extended Data Figure 10, no such dependence exists in the 300 period SL with 25% interface coverage with ErAs dots. The modulation frequency dependence in the reference samples is indicative of ballistic transport. A lower modulation frequency leads to a deeper penetration of the thermal signal so that the resulting thermal conductivity has fewer trace ballistic effects[49]. In the reference samples, these trace ballistic effects, which affect long-wavelength phonons, are evident in the decreasing thermal conductivity with decreasing thermal penetration depth, or increasing pump modulation frequency[50]. The absence of these effects in the 300 period SL with 25% ErAs interface coverage, however, indicates that the long MFP phonons contributing to heat conduction in GaAs/AlAs SLs are reduced to nonobservable levels in the SLs with ErAs dots. This observation is consistent with the consequences of localization.

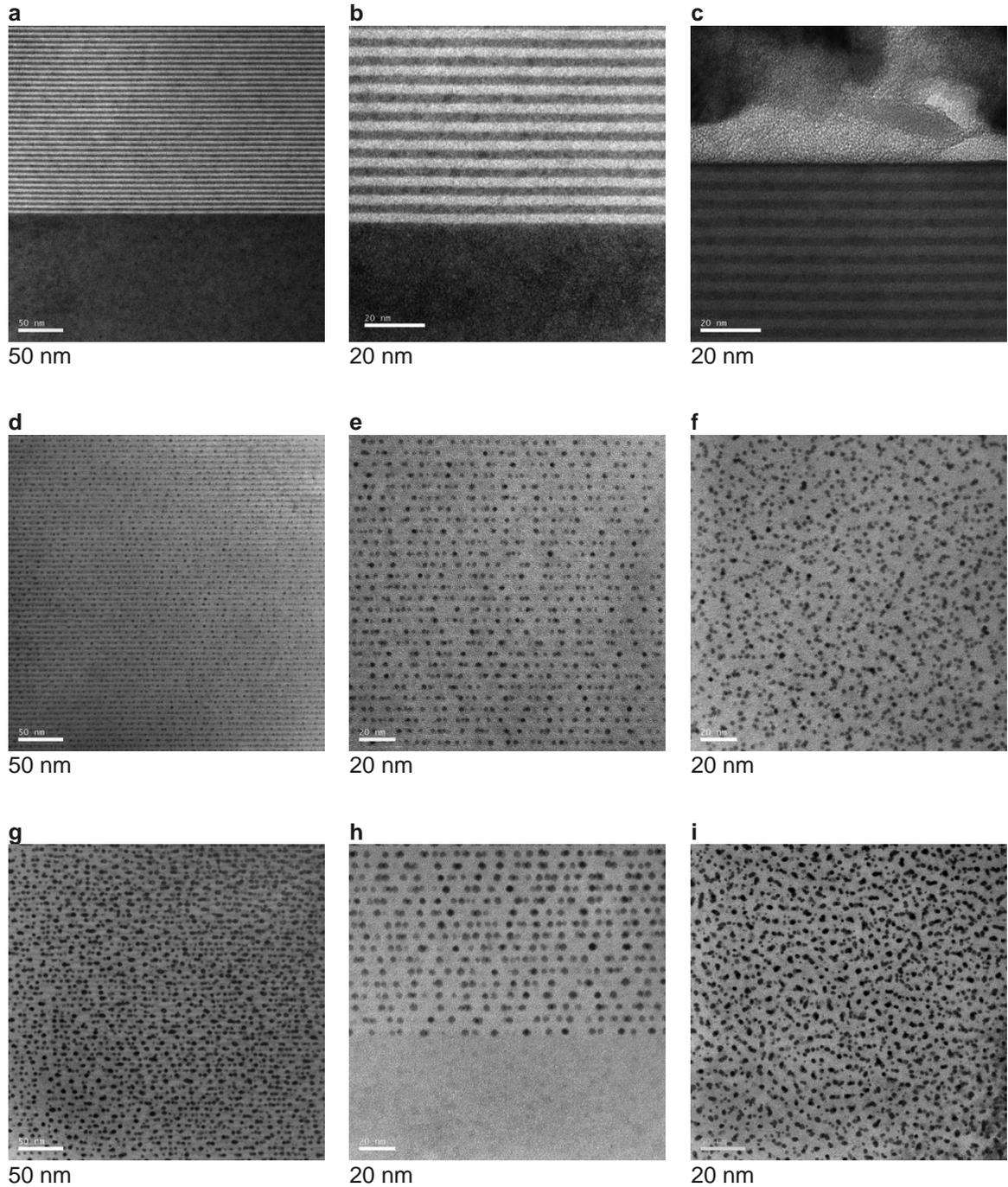

**Extended Data Figure 1 a-c,** 300 period reference GaAs/AlAs SL with no ErAs dots: **a,** cross-sectional TEM, **b,** TEM of the bottom several layers, and **c,** TEM of the top several layers; **d-f,** 300 period GaAs/AlAs SL with 8% areal coverage with ErAs nanodots: **d,** cross-sectional TEM, **e,** close up cross-sectional TEM of the midsection of the SL, and **f,** plan-view TEM; **g-i,** 300 period GaAs/AlAs SL with 25% areal coverage with ErAs nanodots: **g,** cross-sectional TEM, **h,** cross-sectional TEM of the bottom region of the SL, and **i,** plan-view TEM.

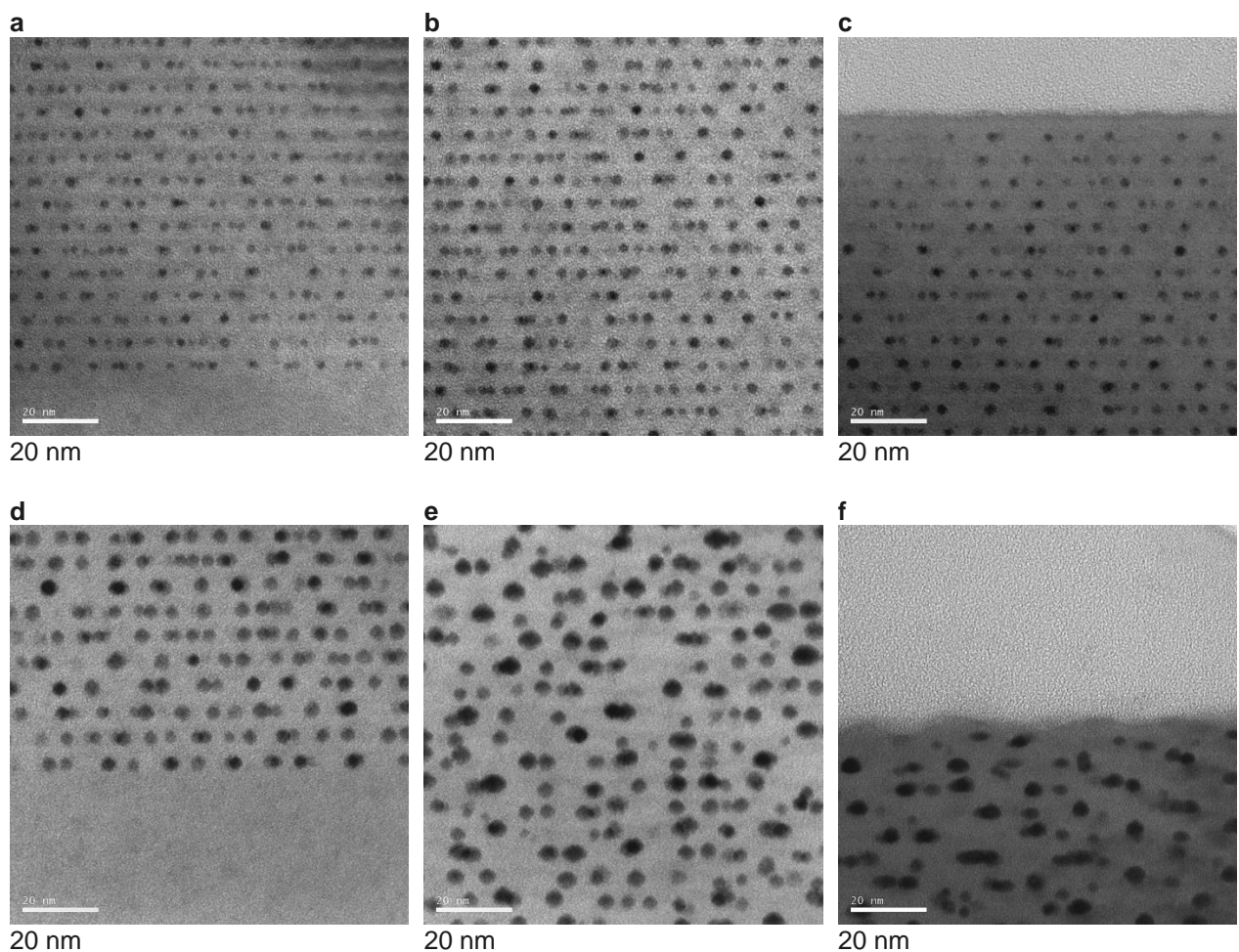

**Extended Data Figure 2** XTEM images of the **a,** bottom, **b,** middle, and **c,** top of the 300-pd SLs with 8% ErAs coverage and the **d,** bottom, **e,** middle, and **f,** top of the 300-pd SLs with 25% ErAs coverage. The samples do not have large defects. Note that the 8% samples show consistent dots distribution from bottom to top, while the 25% samples show slight variations. Localization is observed in both 8% and 25% samples, ruling out that the cause of experimental observations is due to dots variation.

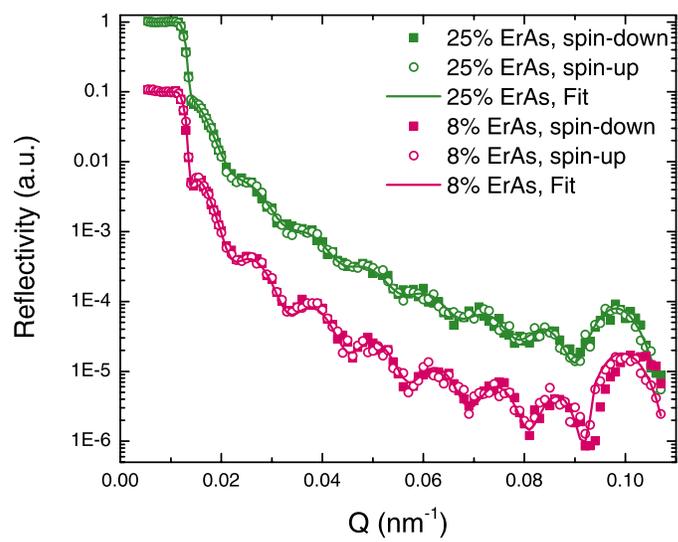

**Extended Data Figure 3** Polarized neutron reflectivity curves of 8-period of 25% ErAs doped (green) and 8% ErAs doped (magenta) GaAs/AlAs SL, measured at $\mu_0 H = 0.7$T and T=4K. The 8% curve is vertically shifted for visual clarity.

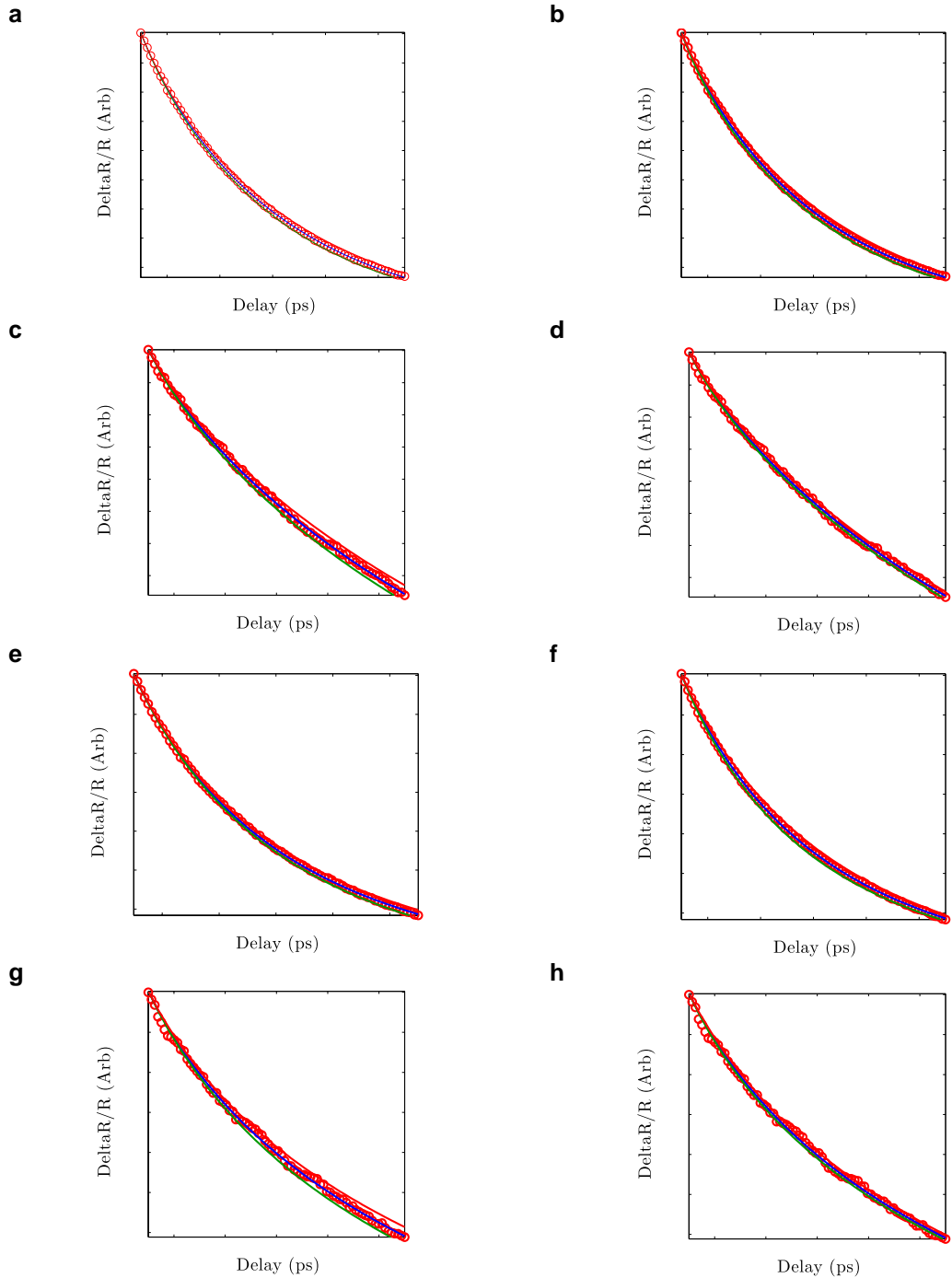

**Extended Data Figure 4** Representative TDTR data (open red circles) sets and fits for **a**-**d**, 16 and **e**-**h**, 300 period samples with no ErAs coverage. The four data sets, **a, b, e,** and **f,** were taken at 40 K and the data sets in **c, d, g,** and **h,** were taken at 296 K. The plots on the left **a, c, e,** and **g,** show the sensitivities of the fitting to the thermal conductivity of the SLs while the plots on the right **b, d, f,** and **h,** show the sensitivity of the fitting to the interface conductance between the Al and the SL. +/- 10% Lines of best fit are in red and green.

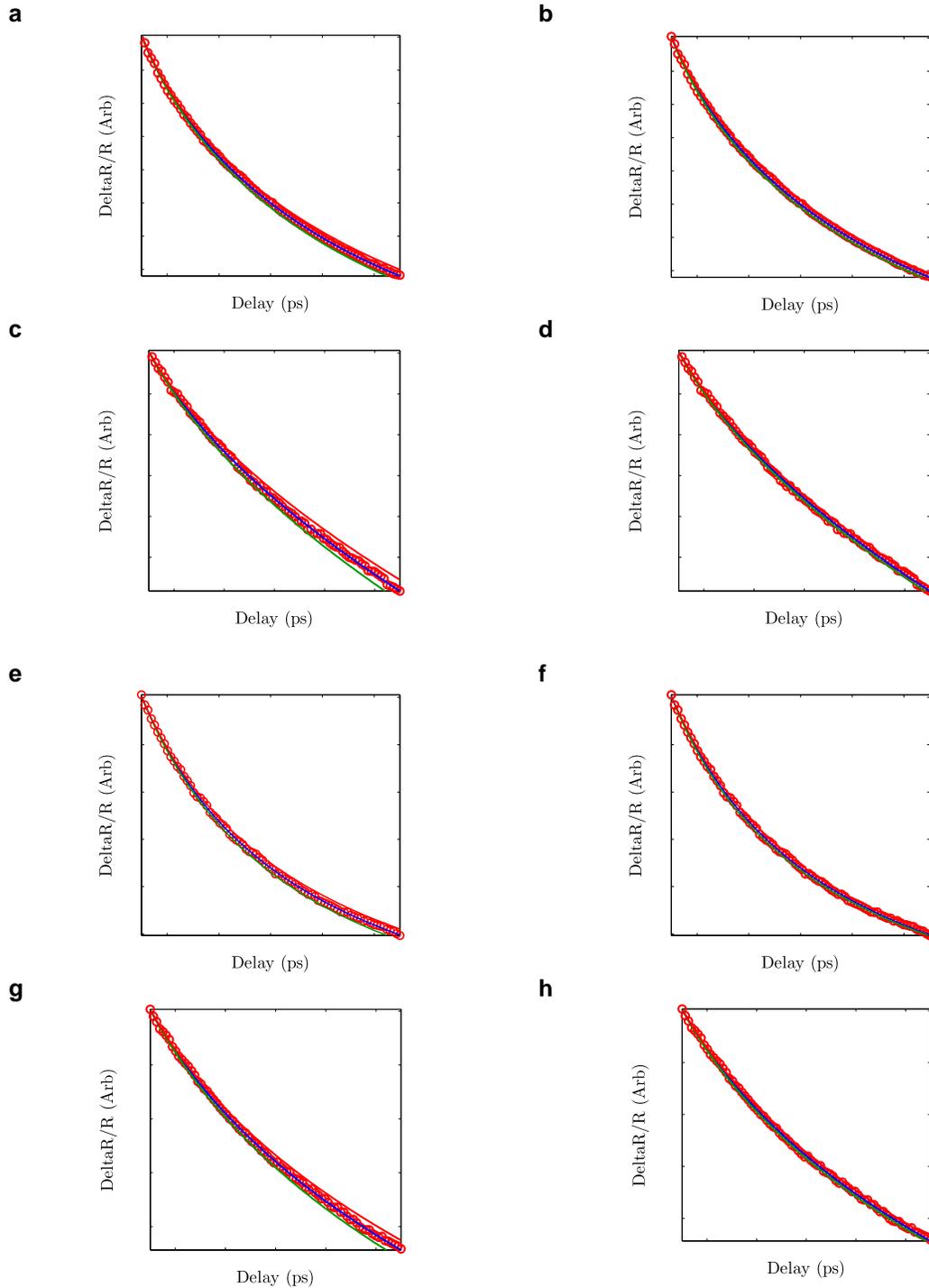

**Extended Data Figure 5** As in Extended Data Figure 4, representative TDTR data (open red circles) sets and fits for **a-d,** 16 and **e-h** 300 period samples with 25% of the interface covered in ErAs nanodots. The four data sets, **a, b, e,** and **f,** were taken at 40 K and the data sets in **c, d, g,** and **h,** were taken at 296 K. The plots on the left **a, c, e,** and **g,** show the sensitivities of the fitting to the thermal conductivity of the SLs while the plots on the right **b, d, f,** and **h,** show the sensitivity of the fitting to the interface conductance between the Al and the SL, G. +/- 10% Lines of best fit are in red and green.

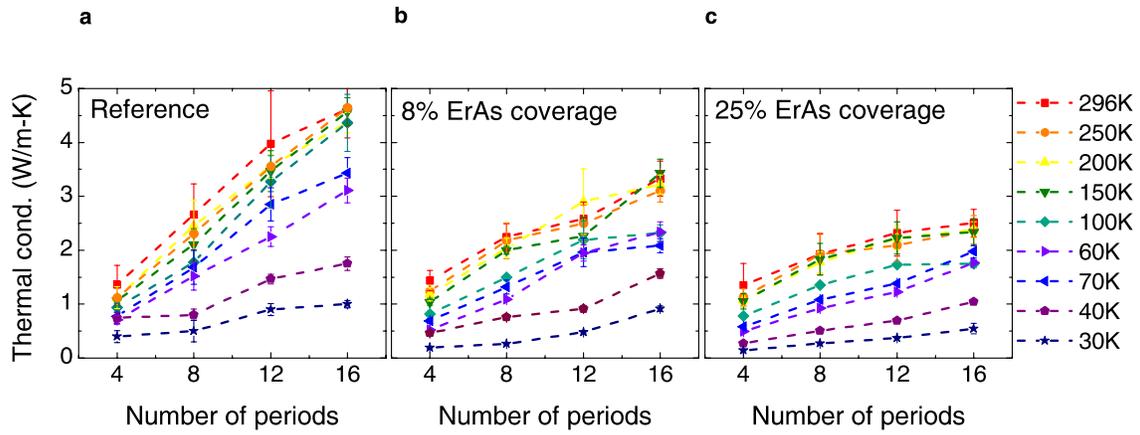

**Extended Data Figure 6** SL thermal conductivities for short period SLs as a function of the number of periods and at different temperatures for **a**, reference GaAs/AlAs SLs with no ErAs nanodots, **b,** GaAs/AlAs SLs with 8% areal coverage with ErAs nanodots, and **c,** GaAs/AlAs SLs with 25% areal coverage with ErAs nanodots.

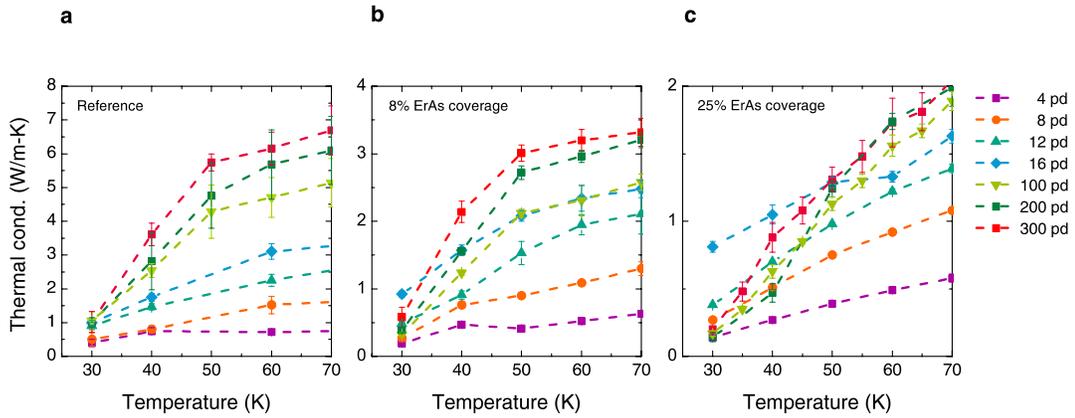

**Extended Data Figure 7** Low-temperature SL thermal conductivities for **a**, reference GaAs/AlAs SLs with no ErAs nanodots, **b,** GaAs/AlAs SLs with 8% areal coverage with ErAs nanodots, and **c,** GaAs/AlAs SLs with 25% areal coverage with ErAs nanodots.

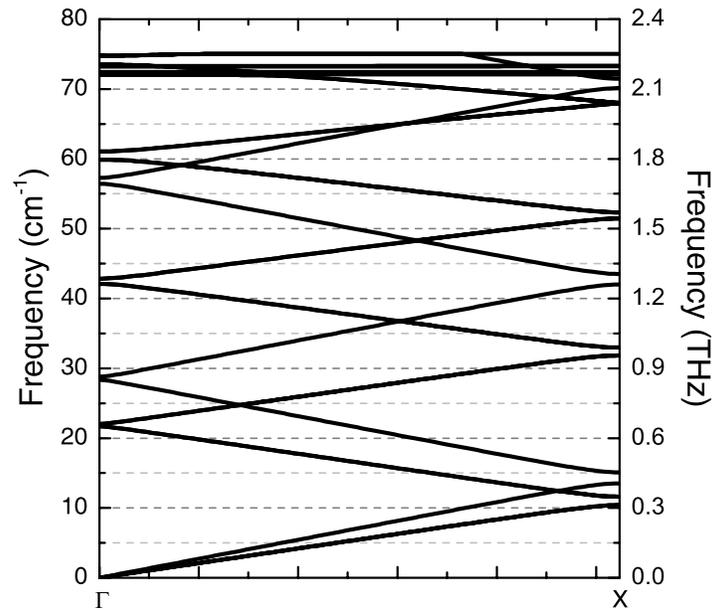

**Extended Data Figure 8** Theoretical SL phonon dispersion relation ω(*k*) from Γ to X.

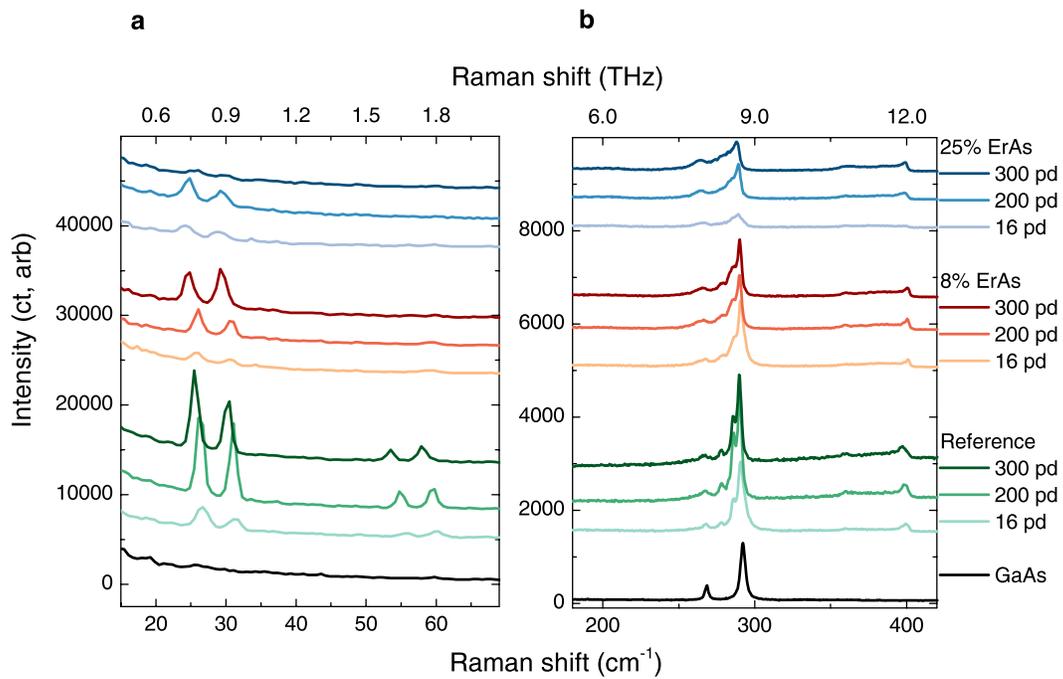

**Extended Data Figure 9** Raman spectra for representative SLs (16, 200, and 300 period SLs with each level of ErAs coverage) in both the **a,** low-frequency (15-70 cm$^{-1}$) and **b,** high-frequency regimes (200-400 cm$^{-1}$).

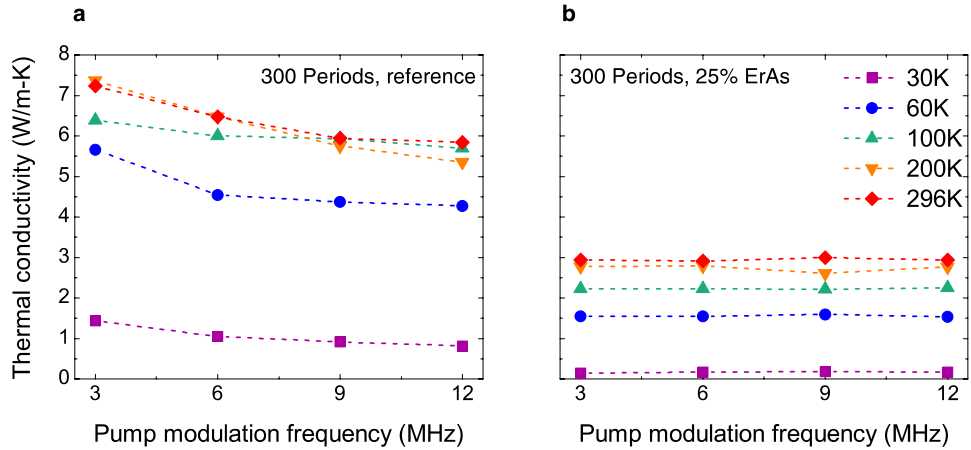

**Extended Data Figure 10** Thermal conductivity as a function of pump modulation frequency at a range of temperatures for the samples with **a,** no ErAs at the interfaces and the samples with **b,** 25% of the interfaces covered with ErAs nanodots. There is a clear trend of decreasing thermal conductivity with increasing modulation frequency in the samples with no ErAs, whereas there is no modulation frequency dependence of the thermal conductivity in the samples with 25% ErAs coverage.

**Author Contributions:**

M. N. Luckyanova performed the TDTR experiments. J. Mendoza conducted Green's functions calculations. H. Lu and A. Gossard fabricated the samples used in the experiments. S. Huang, A. A. Puretzky, and M. S. Dresselhaus conducted Raman measurements. J. Zhou performed first principles calculations. M. Li, A. J. Grutter, B. J. Kirby, and M. N. Luckyanova conducted neutron reflectometry experiments. G. Chen supervised the research and contributed to data interpretation. M. N. Luckyanova and G. Chen wrote the paper.